\documentclass{article}
\usepackage{ijcai22}

\usepackage{times}
\usepackage{soul}
\usepackage{url}
\usepackage[hidelinks]{hyperref}
\usepackage[utf8]{inputenc}
\usepackage[small]{caption}
\usepackage{graphicx}
\usepackage{amsmath}
\usepackage{amsthm}
\usepackage{booktabs}
\usepackage{animate}
\usepackage{algorithm}
\usepackage{algpseudocode}

\usepackage{bm}
\usepackage{multirow}
\usepackage{bbold}
\usepackage{empheq}
\newcommand\myeq{\stackrel{\mathclap{\normalfont\mbox{def}}}{=}}
\title{Music-to-Dance Generation with Optimal Transport}
\author{
Shuang Wu$^{1,2}$
\and
Shijian Lu$^2$\thanks{Corresponding Author}\And
Li Cheng$^3$
\affiliations
$1$Black Sesame Technologies \quad
$^2$Nanyang Technological University \quad
$^3$University of Alberta
\emails
wushuang@outlook.sg,
shijian.lu@ntu.edu.sg,
lcheng5@ualberta.ca
}

\begin{document}

\maketitle

\begin{abstract}
Dance choreography for a piece of music is a challenging task, having to be creative in presenting distinctive stylistic dance elements while taking into account the musical theme and rhythm. It has been tackled by different approaches such as similarity retrieval, sequence-to-sequence modeling and generative adversarial networks, but their generated dance sequences are often short of motion realism, diversity and music consistency. In this paper, we propose a Music-to-Dance with Optimal Transport Network (MDOT-Net) for learning to generate 3D dance choreographies from music. We introduce an optimal transport distance for evaluating the authenticity of the generated dance distribution and a Gromov-Wasserstein distance to measure the correspondence between the dance distribution and the input music. This gives a well defined and non-divergent training objective that mitigates the limitation of standard GAN training which is frequently plagued with instability and divergent generator loss issues. Extensive experiments demonstrate that our MDOT-Net can synthesize realistic and diverse dances which achieve an organic unity with the input music, reflecting the shared intentionality and matching the rhythmic articulation. Sample results are found at \url{https://www.youtube.com/watch?v=dErfBkrlUO8}.
\end{abstract}

\section{Introduction}
Dance and music are intimately related. They share a movement form wherein for dance, movements are articulated visually as body motion whereas for music, movements manifest themselves via an auditory and allegorical nature. These movements, whether visual in dance or auditory in music, evoke emotions and feelings of intentionality. Looking at this intimate connection between dance and music from an artificial intelligence perspective, an intriguing question would be whether a computational model can generate a coherent and meaningful dance sequence for a given piece of music. This is an ambitious task which is useful for simulation, behavior understanding and ultimately benefit the vast community of dancers and musicians.

\begin{figure}[t]
\begin{center}
   \includegraphics[width=1\linewidth]{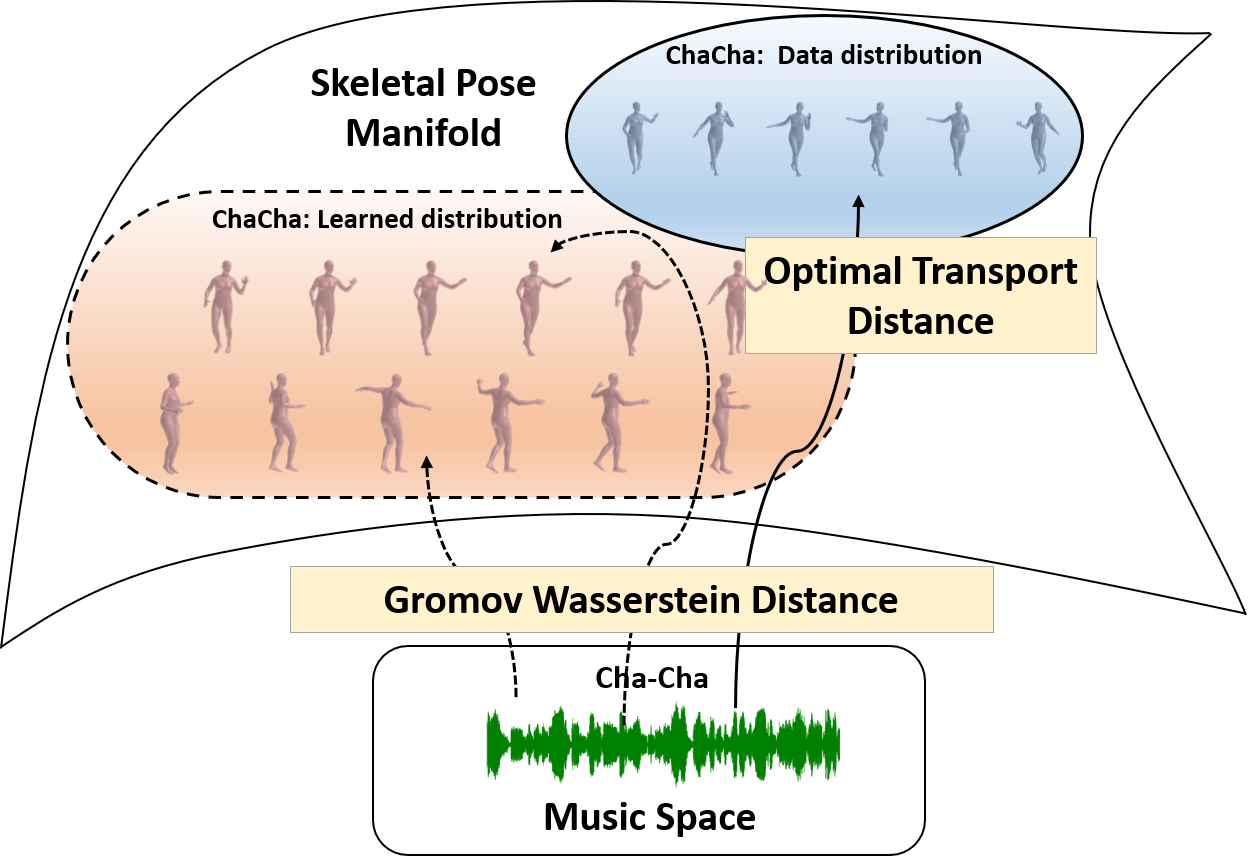}
\end{center}
   \caption{The proposed MDOT-Net generates a matching dance sequence given a music input. We model different genres of music and dance styles (e.g. cha cha and tango) by distributions lying within \emph{Music Spaces} and an \emph{Articulated Pose Manifold}, where the authenticity of the generated dance distributions is evaluated by an \emph{Optimal Transport Distance} and the harmony between the music and the generated dance is measured by a \emph{Gromov-Wasserstein Distance}.}
   \label{fig:problem}
\end{figure}

In this paper, we present a framework generating 3D dance choreographies from music sequences. There are multiple challenges to be addressed for this task.
1) The generated dance motions have to be realistic and adhere to the idiosyncratic distinctions of the dance style. For example, a generated waltz sequence should reflect stylistic elements that are recognizable (even to the non-expert observer).
2) Dance choreography is inherently diverse and multiple choreographic interpretations for the same musical piece are ubiquitous. An adequate computational model would have to generate diverse and multimodal dance kinematics.
3) Choreographing for a piece of music has to take into consideration the rhythmic articulation, melody, theme and variation to achieve an organic unity, a challenging feat even for a professional human expert. The generated dance should be intricately bonded with the music input through a shared intentionality of the movement form, and reflect the melodic styles and rhythmic articulation.

Earlier works typically adopted a similarity retrieval approach \cite{shiratori2006dancing,fan2011example,lee2013music} which lacks creativity. The sequence to sequence modeling approach in \cite{tang2018dance} is also limited to a single output and unable to generate diverse dances for any piece of input music. A recent class of works turned to Generative Adversarial Networks (GANs) \cite{lee2019dancing,ren2020self} for multimodal generation and enhance diversity. However, both \cite{lee2019dancing,ren2020self} focused on 2D choreographies which lack in the dynamic richness and pose realism of 3D choreographies. Furthermore, tuning the adversarial training schemes in these GAN approaches is an arduous task \cite{artetxe2018robust}. Discriminating the music and dance correspondence by mapping to a common embedding space also tends to be inadequate, resulting in lack of coherence between the dance and music.

To address these challenges and overcome the shortcomings of existing methods, we leverage optimal transport (OT) theory \cite{villani2008optimal} and propose a Music-to-Dance with Optimal Transport Network (MDOT-Net). As illustrated in Figure~\ref{fig:problem}, for a given music input, MDOT-Net generates diverse dance sequences that correspond to trajectories over a articulated pose manifold. Directly working with dance distributions supported on the articulated pose manifold is advantageous for realism of the dance poses, allowing subtle stylistic distinctions and nuances to be reflected. We evaluate the optimal transport distance between the generated and data distribution on this manifold. This offers several advantages such as the capability to handle non-overlapping distributions (a major issue for Jensen-Shannon divergence) \cite{arjovsky2017wasserstein} and non-divergent generator loss by reframing the adversarial training as an optimization problem \cite{genevay2018learning,salimans2018improving}.

Since dance and music exist in different domains, it is difficult to quantitatively gauge their differences directly. Therefore, we propose a Gromov-Wasserstein distance \cite{memoli2011gromov,bunne2019learning} to compare distributions over different domains (music space and articulated pose manifold). The Gromov-Wasserstein distance compares distributions in \emph{relational} terms. The intuition is that for a matching pair of dance and music $(D,M)$, a generated dance sequence $D'$ is likely a good match for music sequence $M'$ if $D'$ is close to $D$ and $M'$ is close to $M$. Similar to the optimal transport distance in facilitating adversarial training, the Gromov-Wasserstein distance enables a more efficient approach in assessing the music and dance correspondence.

Our contribution are summarized as follows. 1) We develop a novel optimal transport framework for music to dance generation. The authenticity of the generated dance is measured via an optimal transport distance on the manifold of articulated poses. 2) A Gromov-Wasserstein distance is incorporated to facilitate learning cross modal generations from the music space to the articulated pose manifold through a \emph{relational} rather than absolute measure of the music and dance similarity. 3) Our MDOT-Net can generate realistic and diverse 3D dance sequences faithful to the rhythm and melody of a given music input.

\section{Related Works}
\textbf{Optimal Transport for Generative modeling} \quad
Optimal transport \cite{villani2008optimal} defines a metric distance for probability distributions over arbitrary spaces. The generative modeling problem is reframed as finding an optimized transport for aligning the model distribution and data distribution. However, solving the optimal transport problem is expensive, and this computational burden presented major hurdles for employing optimal transport for generative modeling. The Wasserstein GAN \cite{arjovsky2017wasserstein} turned to the dual optimal transport problem and proposed a discriminator approximating 1-Lipschitz functions for GAN training. An alternative line of work was pursued in \cite{cuturi2013sinkhorn,genevay2016stochastic}, in which the introduction of an entropic regularization term reduces the computational cost. The regularized primal optimal transport problem is amenable to backpropagation training \cite{genevay2018learning,salimans2018improving}.

\cite{memoli2011gromov} generalizes optimal transport for comparing distributions supported on different spaces, introducing the Gromov-Wasserstein distance as a notion of distance between intra-domain distances. The Gromov-Wasserstein distance is a promising metric for learning cross-domain correspondences \cite{bunne2019learning} such as unsupervised language translation \cite{alvarez2018gromov} or graph matching \cite{xu2019gromov}.

\textbf{Dance Generation} \quad
Earlier works generally utilised similarity retrieval \cite{shiratori2006dancing,fan2011example,lee2013music}. A major drawback is that the synthesized choreography appears rigid and lacks creativity, simply arranging the dance moves in the training data with unnatural transitions. \cite{tang2018dance} employs Long Short Term Memories (LSTM) in a sequence to sequence modeling framework that generates motion features from encoded musical features. However, in using a L2 loss for the dance sequences, the synthesized motion are unrealistic and tends to incur motion freezing for longer sequences. Another shortcoming is the inability to generate diverse dance sequences.

\cite{huang2021dance} proposes curriculum learning with L1 loss on the dance sequences to alleviate the motion freezing issue. It also introduces a noise vector on top of the encoded musical feature to enable multimodal generation. An alternative approach utilises GANs to synthesize multimodal dances. \cite{ren2020self} adopts a GAN framework with both a local and global discriminator to measure discrepancies between dance sequences. \cite{lee2019dancing} proposes a two-phase framework by first learning a decomposition of dance sequences into basic dance motion units and subsequently composing these dance units into a dance sequence with a GAN. These works focused on 2D poses, losing the geometric richness and realism of 3D motion. Important cues such as the dancer's position cannot be clearly put into perspective and invariance of bone lengths across frames is not enforced. More recently, \cite{valle2021transflower,li2021ai,wu2021dual} explored transformer-based models for music-to-dance generation with \cite{wu2021dual} proposing a dual learning framework of concurrently learning music composition conditional on dance inputs.

\begin{figure*}[ht]
\begin{center}
   \includegraphics[width=1.00\linewidth]{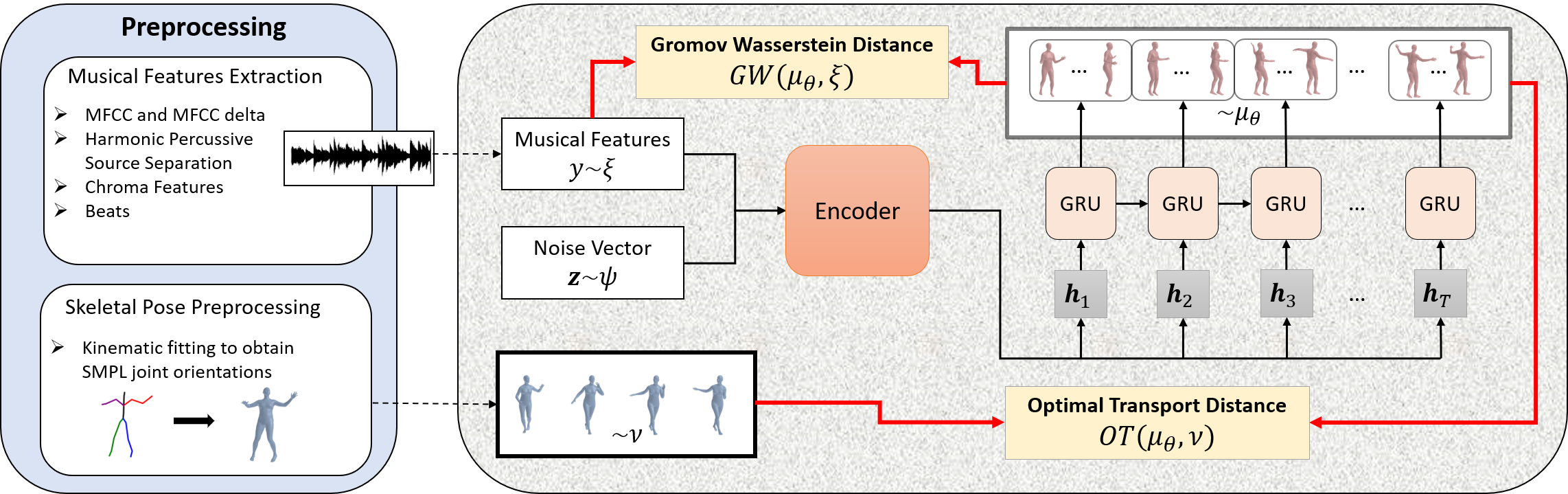}
\end{center}
   \caption{Overview: An \emph{Encoder} maps the extracted \emph{Musical Features} and \emph{Noise Vector} into latent vectors $\mathbf{h}_1,\cdots,\mathbf{h}_N$. Each $\mathbf{h}_i$ corresponds to an inter-beat sequence, comprising a hierarchical representation of global musical feature and local beat-level features. A \emph{GRU decoder} generates a dance sequence from $\mathbf{h}_i$ for each beat event. The authenticity of the generated sequences is measured by an \emph{Optimal Transport Distance}, and the music-to-dance matching is measured by a \emph{Gromov-Wasserstein Distance}.}
\label{fig:framework}
\end{figure*}

\section{Our Approach}
\noindent\textbf{Music Input Preprocessing} \quad
The music waveform is sampled at 48kHz. We do not use the raw waveform as input as it would be too computationally expensive with too much redundancies. Following existing works \cite{tang2018dance,ren2020self}, we adopt a similar procedure of extracting Mel-frequency cepstral coefficients (MFCC) and MFCC delta features which constitute low level sound features \cite{muller2015fundamentals}. To incorporate additional high level musical information, we further extract chroma features which correspond to pitch and melody as well as beats which relate to rhythm.

\noindent\textbf{Pose Preprocessing}\quad
We first perform an inverse kinematics fitting to obtain the joint orientation parameters for each dance pose in the Skinned Multi-Person Linear (SMPL) model \cite{SMPL:2015}. Key advantages of this over 3D positions representation include: 1) bone length invariance and rotational degrees of freedom are inherently inbuilt in this manifold; 2) We can easily normalize the bone lengths across performers and the global orientation of a dance sequence.

\subsection{Formal Problem Statement}
We denote the dance distribution as $\nu$ and the music distribution as $\xi$. As illustrated in Figure~\ref{fig:framework}, the generator with parameters $\theta$ learns a parametric mapping $g_{\theta}$ that maps an input music sequence $y$ (sampled from $\xi$) and noise vector $z$ (sampled from a Gaussian distribution $\psi$) to a generated dance sequence $\tilde{x}$. This gives the model distribution of generated dances $\mu_{\theta}$\footnote{Formally, $\mu_{\theta}$ is the push forward probability distribution of $(\xi,\psi)$ under the generator mapping $g_{\theta}$, \emph{i.e.} $\mu_\theta\myeq{g_{\theta_{\#}}(\xi,\psi)}$.}. A crucial aspect of our MDOT-Net is in introducing an optimal transport distance $OT(\mu_{\theta},\nu)$ and a Gromov-Wasserstein distance $GW(\mu_{\theta},\xi)$ as objective functions to facilitate learning of the generator parameters $\theta$
\begin{equation}\label{eqn:problem_statement}
\begin{aligned}
&\arg\min_{\theta} \;OT (\mu_{\theta},\nu) + GW (\mu_{\theta},\xi).
\end{aligned}
\end{equation}
In what follows, we present the definitions for $OT (\mu_{\theta},\nu)$ and $GW (\mu_{\theta},\xi)$ and algorithms for computing them.

\subsection{Optimal Transport Distance} 
For the generated dance distribution $\mu_{\theta}$ and the data dance distribution $\nu$, we consider each as a discrete distribution with $m$ samples. We have $\mu_{\theta}=\frac{1}{m}\sum_{i=1}^m\delta_{\tilde{x}_i}$, $\nu=\frac{1}{m}\sum_{j=1}^m\delta_{x_j}$ where $\delta$ denotes the Dirac delta distribution. The optimal transport distance between $\mu_{\theta}$ and $\nu$ is obtained via the following optimal transport problem \cite{kantorovitch1958translocation}:
\begin{equation}\label{eqn:ot_discrete}
OT_c (\mu_{\theta},\nu)=\min_{\gamma \in \Gamma} \sum_{i,j} \gamma_{ij}c(\tilde{x}_i,x_j).
\end{equation}
Intuitively, the optimal transport distance is the minimum total cost of matching $\mu_{\theta}$ with $\nu$ with $c(\tilde{x},x)$ denoting the unit cost of moving generated sequence $\tilde{x}$ to data sequence $x$. This is optimized over the set $\Gamma$ of all possible transport plans:
\begin{equation} \label{eqn:transport_plan}
\Gamma=\left\{\gamma\in\mathbb{R}^{m\times m}_+\mid\forall{i}\sum_{j}\gamma_{ij}=1,\forall{j}\sum_{i}\gamma_{ij}=1 \right\}.
\end{equation}

\noindent\textbf{Cost function $c$} \quad
A $T$-frames dance sequence is given by $x=(\mathbf{p}_1,\cdots,\mathbf{p}_T)\in\mathcal{X}^T$ where each pose comprises of joint rotations, \emph{i.e.} $\mathbf{p}_i=(R_{i,1},\cdots,R_{i,J})$ \footnote{We disregard global translation of the pose.}. We propose a cost function measuring the squared geodesic distance on the $SO(3)$ rotation manifold:
\begin{equation}\label{eqn:rotation_cost}
\begin{aligned}
c_R(\tilde{x},x) &= \sum_{i=1}^T \sum_{j=1}^J \text{geodesic}_{SO(3)}(\tilde{R}_{i,j},R_{i,j})^2\\
&=\sum_{i=1}^T \sum_{j=1}^J \left\lvert \arccos \left[\frac{\text{Tr}(\tilde{R}_{i,j}^TR_{i,j})-1}{2}\right]
\right\rvert^2.
\end{aligned}
\end{equation}

\noindent\textbf{Solving for Equation~\ref{eqn:ot_discrete}} \quad
For distributions supported on an Euclidean space $\mathbb{R}^d$ with L1 cost function $c(\tilde{x},x)=|\tilde{x}-x|$, the optimal transport distance is known as the Wasserstein distance. The can be solved through 1-Lipschitz functions in the dual formulation as proposed in Wasserstein GAN \cite{arjovsky2017wasserstein}. Our case is more complicated in that the distributions are over a manifold with a squared geodesic cost function. To solve for this, we follow \cite{genevay2016stochastic} in introducing a regularization term \begin{equation}\label{eqn:ot_regularized}
\begin{aligned}
OT_{c,\epsilon}(\mu_{\theta},\nu) =\min_{\gamma \in \Gamma}\sum_{i=1}^n\sum_{j=1}^n \gamma_{ij}c(\tilde{x}_i,x_j)+\epsilon I(\gamma)
\end{aligned}
\end{equation}
where $I(\gamma) = \sum_{i,j}\gamma_{ij}\log_2\gamma_{ij}$ is the mutual information of $\mu_{\theta},\nu$.
This regularization transforms the primal Kantorovich problem into a convex optimization problem and Eqn~\ref{eqn:ot_regularized} admits a solution of the form $\gamma*=\text{diag}(\mathbf{a})K \text{diag}(\mathbf{b})$ where $K_{ij}\myeq{-\exp(c(\tilde{x}_i,x_j)/\epsilon})$. We compute the optimal transport distance with Algorithm~\ref{algo:OT}. This algorithm has the crucial advantage of being amenable to backpropagation \cite{genevay2018learning} and faster convergence.
\begin{algorithm}
\caption{OT Distance for batch of $m$ samples with Sinkhorn-Knopp algorithm}
\label{algo:OT}
\begin{algorithmic}
\renewcommand{\algorithmicrequire}{\textbf{Input:}}
\Require gen. dance sequences $\widetilde{\mathbf{X}}=\{\tilde{x}_i\}_{i=1}^m$
\Require data dance sequences $\mathbf{X}=\{x_j\}_{j=1}^m$
\renewcommand{\algorithmicrequire}{\textbf{Hyperparameters:}}
\Require regularization $\epsilon$, Sinkhorn iterations $L$
\State Dance Cost Matrix  $C_{ij}=c_R(\tilde{x}_i,x_j)$ from  Eqn~\ref{eqn:rotation_cost},
\State $K = \exp(-C/\epsilon)$
\State $\mathbf{b}^{(0)}=\mathbb{1}_m$ where $\mathbb{1}_m=(1,\cdots,1)^T\in\mathbb{R}^m$ 
\For{$\ell = 1:L$}
\State $\mathbf{a}^{(\ell)} = \mathbb{1}_m \oslash K\mathbf{b}^{(\ell-1)}$, $\mathbf{b}^{(\ell)} = \mathbb{1}_m \oslash K^T\mathbf{a}^{(\ell)}$
\State $\oslash$ denotes component-wise division
\EndFor
\Ensure $OT_{\epsilon} (\widetilde{\mathbf{X}},\mathbf{X}) = \sum_{i,j}C_{ij}a_{i}^{(L)}K_{ij}b_{j}^{(L)}$
\end{algorithmic}
\end{algorithm}

Following \cite{salimans2018improving}, we sample two independent mini-batches of data and generated sequence pairs $(\widetilde{\mathbf{X}}, \mathbf{X}), (\widetilde{\mathbf{X}}', \mathbf{X}')$ in order to compute the following unbiased optimal transport distance $\overline{OT}_{\epsilon}$
\begin{equation}
\small
\begin{aligned}
\overline{OT}_{\epsilon}&=OT_{\epsilon}(\widetilde{\mathbf{X}},\mathbf{X})+OT_{\epsilon}(\widetilde{\mathbf{X}}',\mathbf{X})+OT_{\epsilon}(\widetilde{\mathbf{X}},\mathbf{X}')\\
&+OT_{\epsilon}(\widetilde{\mathbf{X}}',\mathbf{X}')-2OT_{\epsilon}(\widetilde{\mathbf{X}},\widetilde{\mathbf{X}}')-2OT_{\epsilon}(\mathbf{X},\mathbf{X}'). \label{eqn:ot_unbiased}
\end{aligned}
\end{equation}

\subsection{Gromov-Wasserstein distance}
Comparing the similarity of dance distribution $\mu_{\theta}$ and music distribution $\xi$ invokes a cross domain learning problem. The Gromov-Wasserstein distance is defined as a \emph{relational} distance between the respective costs within each distribution. The cost function for the dance pose manifold is defined in Eqn~\ref{eqn:rotation_cost}. For music distributions, we learn an embedding $f$ and define the cost as the L1 distance in this embedding space
\begin{equation} \label{eqn:music_cost}
\begin{aligned}
d(y_i,y_j) = \lVert f(y_i)-f(y_j)\rVert_1.
\end{aligned}
\end{equation}
The Gromov-Wasserstein distance for our task is given by
\begin{flalign} \label{eqn:Gromov-Wasserstein}
&\Pi=\left\{\pi\in\mathbb{R}^{m\times m}_+\mid\forall{i}\sum_{j}\pi_{ij}=1,\forall{j}\sum_{i}\pi_{ij}=1 \right\}\\
&GW(\mu_{\theta},\xi)=\min_{\pi\in\Pi}\sum_{i,j,k,l}\lvert c_R(\tilde{x}_i, \tilde{x}_k) - d(y_j, y_l) \rvert^2 \pi_{ij} \pi_{kl}. \nonumber
\end{flalign}

\noindent\textbf{Solving for Eqn~\ref{eqn:Gromov-Wasserstein}} \quad
This may be solved via entropic regularization and projected gradient descent as in Algorithm~\ref{algo:GW}.

\begin{algorithm}[H]
\caption{GW Distance for 2 independent batches of $m$ samples}
\label{algo:GW}
\begin{algorithmic}
\renewcommand{\algorithmicrequire}{\textbf{Input:}}
\Require gen. dance sequences $\widetilde{\mathbf{X}}=\{\tilde{x}\}_{i=1}^m,\widetilde{\mathbf{X}}'=\{\tilde{x}'\}_{i=1}^m$
\Require music sequences $\mathbf{Y}=\{y\}_{i=1}^m,\mathbf{Y}'=\{y'\}_{i=1}^m$
\renewcommand{\algorithmicrequire}{\textbf{Hyperparameters:}}
\Require regularization $\varepsilon$, projection iterations $M$, Sinkhorn iterations $L$
\renewcommand{\algorithmicrequire}{\textbf{Initialize:}}
\Require $\pi^{(0)}_{ij} = \frac{1}{n} \forall{i,j}$
\State Dance Cost Matrix $C_{ij}=c_R(\tilde{x}_i,\tilde{x}'_j)$ from Eqn~\ref{eqn:rotation_cost}
\State Music Cost Matrix $D_{ij} = d(y_i,y'_j)$ from Eqn~\ref{eqn:music_cost}
\For{$l = 1:M$}
\State $E = \frac{1}{m}D^2 \mathbb{1}_m \mathbb{1}_m^T + \frac{1}{m} \mathbb{1}_m \mathbb{1}_m^T C^2 - 2D \pi^{(l-1)} C^T$
\State $K = \exp(-E/\varepsilon)$
\State $\mathbf{b}^{(0)}  = \mathbb{1}_m$
\For{$\ell = 1:L$}
\State $\mathbf{a}^{(\ell)} = \mathbb{1}_m \oslash K\mathbf{b}^{(\ell-1)}$, $\mathbf{b}^{(\ell)} = \mathbb{1}_m \oslash K^T\mathbf{a}^{(\ell)}$
\EndFor
\State $\pi^{(l)}=\text{diag}(\mathbf{a}^{(L)})K\text{diag}(\mathbf{b}^{(L)})$
\EndFor
\Ensure \small$GW_{\varepsilon}(\widetilde{\mathbf{X}},\widetilde{\mathbf{X}}',\mathbf{Y},\mathbf{Y}') = \displaystyle \sum_{i,j,k,l}\lvert C_{ik}-D_{jl}\rvert^2\pi_{ij}^{(M)}\pi_{kl}^{(M)}$
\end{algorithmic}
\end{algorithm}

\subsection{Algorithmic Pipeline}
\begin{figure}[ht]
\begin{center}
\includegraphics[width=1.00\linewidth]{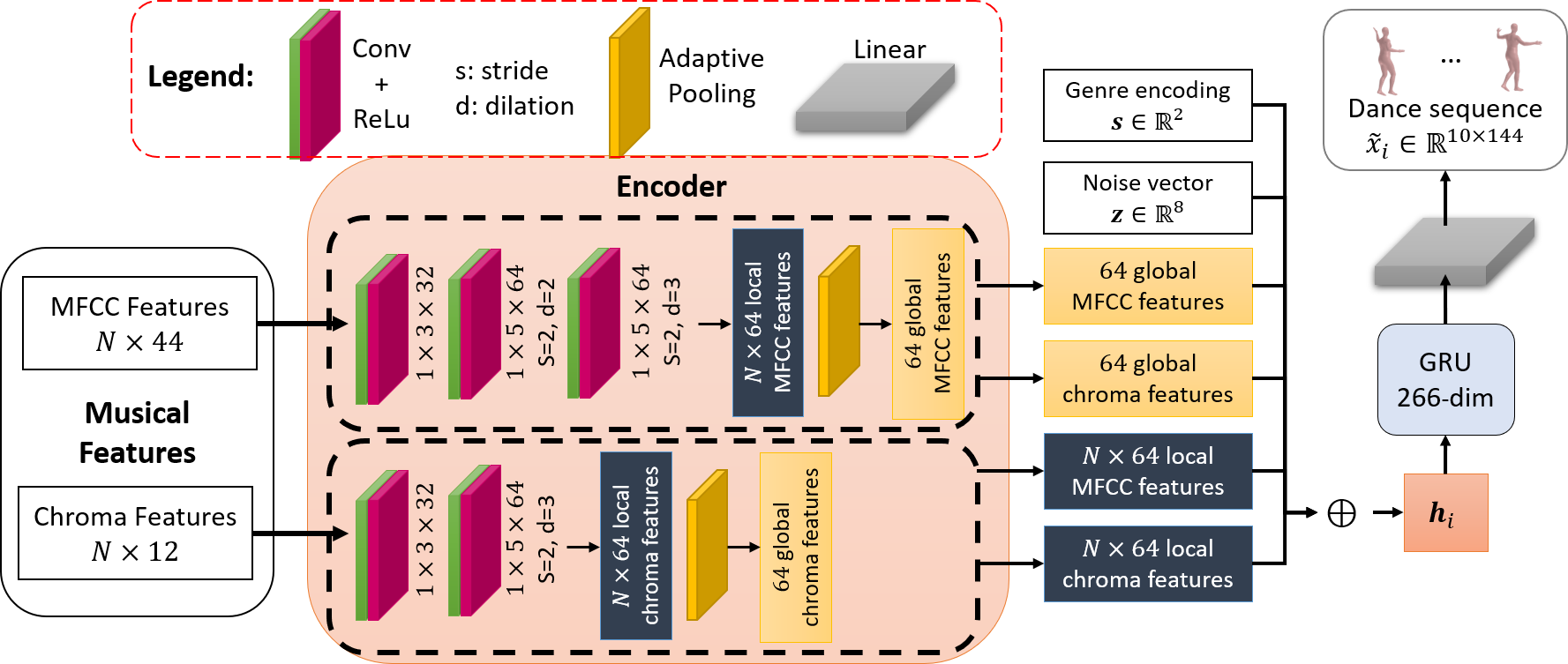}
\end{center}
\caption{Architecture details of encoder and decoder}
\label{fig:encoder}
\end{figure}

An overview of our MDOT-Net is presented in Figure~\ref{fig:framework}. We generate an inter-beat dance sequence directly instead of a frame-by-frame synthesis since this improves the temporal smoothness and also facilitates modeling nuances in dance motions. Our encoder network, as illustrated in Figure~\ref{fig:encoder} processes the musical feature $y$ and a noise vector $\mathbf{z}$ into latent vectors $\mathbf{h}_1,\cdots,\mathbf{h}_T$ where $T$ denotes the total beat events in the music input $y$. Each $\mathbf{h}_i$ consists of a hierarchical representation of the global musical feature and local beat level features. Each $\mathbf{h}_i$ is decoded via a GRU network into a dance sequence $\tilde{x}_i$ consisting of 10 articulated poses. For post-processing, we employ spherical interpolation to fix the frames per second to 25. The training procedure is summarized in Algorithm~\ref{algo:pipeline}. 

Due to space limitations, a discussion on further motivations of our optimal transport based cross-domain sequence-to-sequence learning is presented in the supplementary text.

\begin{algorithm}[H]
\caption{Overall Algorithmic Pipeline}
\label{algo:pipeline}
\begin{algorithmic}
\renewcommand{\algorithmicrequire}{\textbf{Input:}}
\Require music dataset with data distribution $\xi$
\Require dance dataset with data distribution $\nu$
\renewcommand{\algorithmicrequire}{\textbf{Hyperparameters:}}
\Require regularization parameters $\epsilon, \varepsilon$, batch size $m$, learning rate $\alpha$, training epochs $T$, generator to critic update ratio $n_{\text{gen}}$, noise vector distribution $\psi$
\renewcommand{\algorithmicrequire}{\textbf{Initialize:}}
\Require generator parameters $\theta_0$, music embedding $f_0$.
\For{$l = 1:T$}
\State Sample 2 independent mini-batches of music-dance pairs from data and noise vectors from $\psi$, $(\mathbf{X},\mathbf{Y},\mathbf{Z}),(\mathbf{X}',\mathbf{Y}',\mathbf{Z}')$
\State Generate dance sequences as $\widetilde{\mathbf{X}}=g_{\theta}(\mathbf{Y},\mathbf{Z})$ and $\widetilde{\mathbf{X}}'=g_{\theta}(\mathbf{Y}',\mathbf{Z}')$.
\State Compute $\overline{OT}_{\epsilon}$ with Algorithm~\ref{algo:OT} and Equation~\ref{eqn:ot_unbiased}.
\State Compute $GW_{\varepsilon}$ with Algorithm~\ref{algo:GW}.
\If{$l \mod (n_{\text{gen}}) > 0$}
\State $\theta \leftarrow \theta-\alpha\nabla_{\theta}\overline{OT}_{\epsilon}-\alpha\nabla_{\theta}GW_{\varepsilon}$
\Else
\State $f \leftarrow f+\alpha\nabla_{f}GW_{\varepsilon}$
\EndIf
\EndFor
\Ensure Generator network parameters $\theta$
\end{algorithmic}
\end{algorithm}

\section{Experimental Results}
\subsection{Dataset and Implementation Details}
We adopt the public dataset of \cite{tang2018dance} comprising four dance styles, waltz, tango, cha-cha, and rumba. The dances were performed by professional dancers and comprises 3D motion capture data obtained via Vicon MoCap devices. We perform an inverse kinematics fitting to re-parameterize the dance sequence as SMPL \cite{SMPL:2015} parameters, consisting of 3D orientations for 24 joints.

We implement MDOT-Net in PyTorch \cite{paszke2019pytorch}. Additional libraries included Librosa \cite{mcfee2015librosa} for music processing and the Python Optimal Transport Toolbox \cite{flamary2017pot}. The hyperparameters are as follows: regularization parameters are set to $\epsilon=0.1$, $\varepsilon=0.5$; mini-batch size is set to $m=128$; learning rate is set to $\alpha=0.001$; number of Sinkhorn iterations is set to $L=30$; projection iterations is set to $M=20$; generator to critic update ratio is set to $n_{\text{gen}}=10$. The RMSprop optimizer is used. Convergence occurs around 150 epochs.

\subsection{Baselines and Ablation Studies}
\noindent\textbf{Baselines}: As a relatively novel task, few methods have been developed for generating 3D dance sequences from music. Existing methods \cite{ren2020self,huang2021dance} focus on 2D generation, making it difficult to visualize the full dynamic richness of dance motions. As such, we adapted them for 3D generation for fair comparison. 

\noindent\textbf{Ablation Studies}: We further perform two ablation studies for validating the efficacy of the optimal transport distance and Gromov-Wasserstein distance.
\textbf{1) WGAN}: Here we replace the optimal transport and Gromov-Wasserstein objectives with WGAN discriminators. The discriminator for dance sequences is adapted from AGCN \cite{shi2019two} (SOTA for modeling skeletal based motion). A second discriminator serves to determine if the dance sequence matches input music.
\textbf{2) Remove GW}: We remove the Gromov-Wasserstein objective to investigate its effectiveness in establishing music and dance correspondence.

\begin{figure*}[ht]
\centering
\resizebox{\textwidth}{!}
{
\begin{tabular}{|l||c||c|}
\hline
Method & Click $\downarrow$ & Tango dance sequences shown at 1 second intervals
\\ \hline
Ground Truth &
\animategraphics[height=0.92cm]{5}{Figures/GT/}{0}{39} & \includegraphics[height=0.92cm]{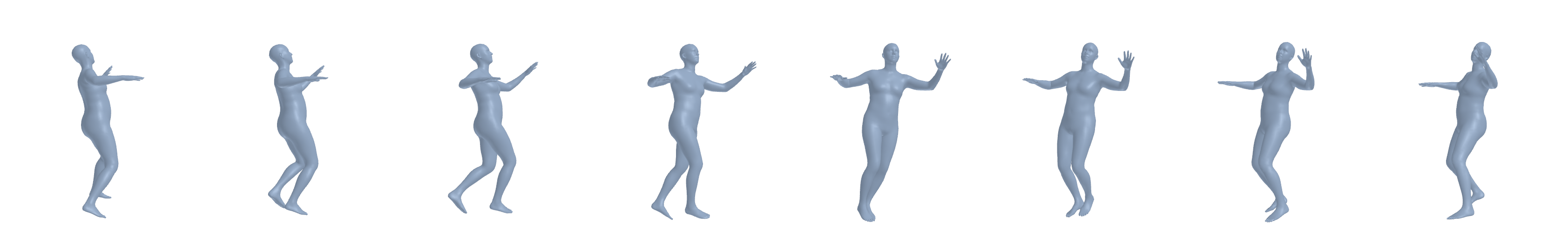}
\\ \hline
\cite{tang2018dance} &
\animategraphics[height=0.92cm]{5}{Figures/Tang2018/}{0}{39} & \includegraphics[height=0.92cm]{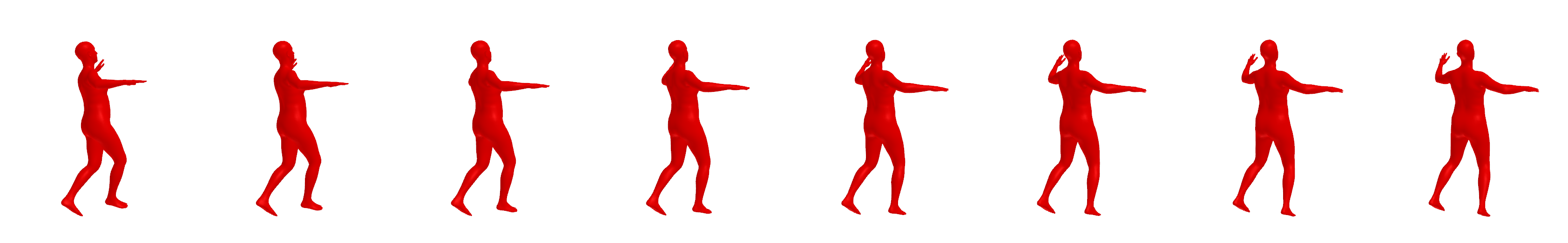}
\\ \hline
\cite{ren2020self} &
\animategraphics[height=0.92cm]{5}{Figures/Ren2020/}{0}{39} & \includegraphics[height=0.92cm]{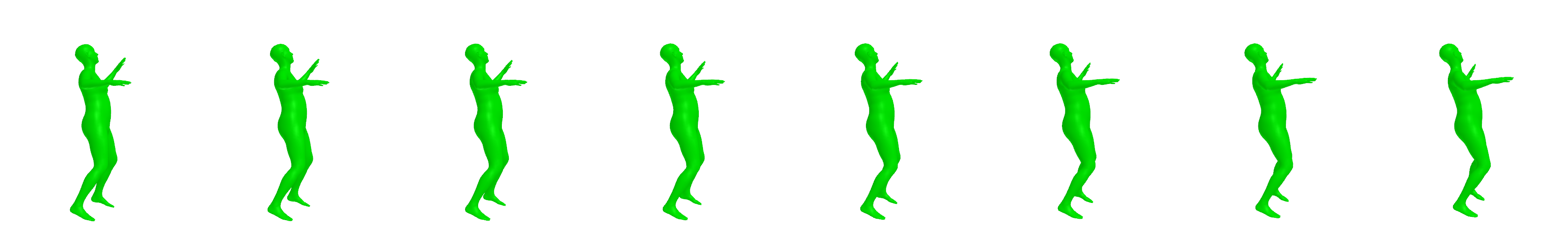}
\\ \hline
\cite{huang2021dance} &
\animategraphics[height=0.92cm]{5}{Figures/Huang2021/}{0}{39} & \includegraphics[height=0.92cm]{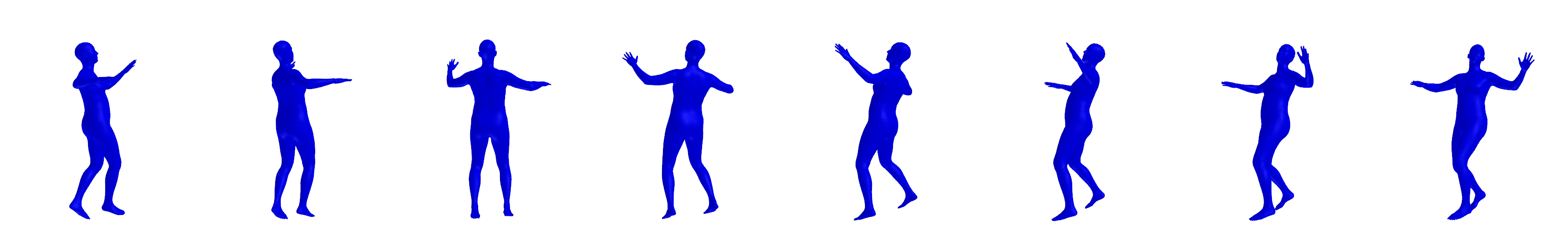}
\\ \hline
Ablation: WGAN &
\animategraphics[height=0.92cm]{5}{Figures/WGAN/}{0}{39} & \includegraphics[height=0.92cm]{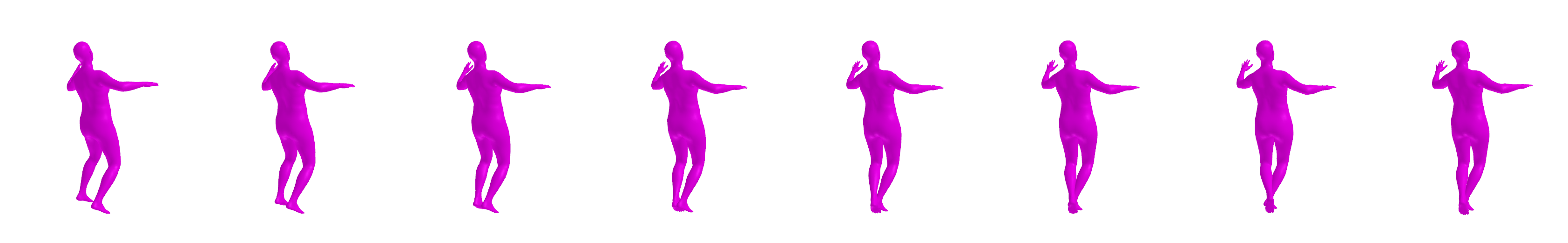}
\\ \hline
Ablation: Remove GW &
\animategraphics[height=0.92cm]{5}{Figures/Remove_GW/}{0}{39} & \includegraphics[height=0.92cm]{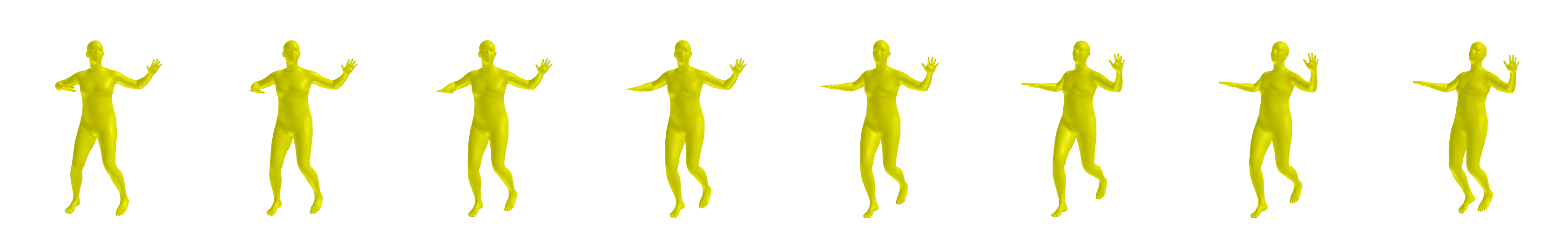}
\\ \hline
MDOT-Net (Ours) &
\animategraphics[height=0.92cm]{5}{Figures/MDOT-Net/}{0}{39} & \includegraphics[height=0.92cm]{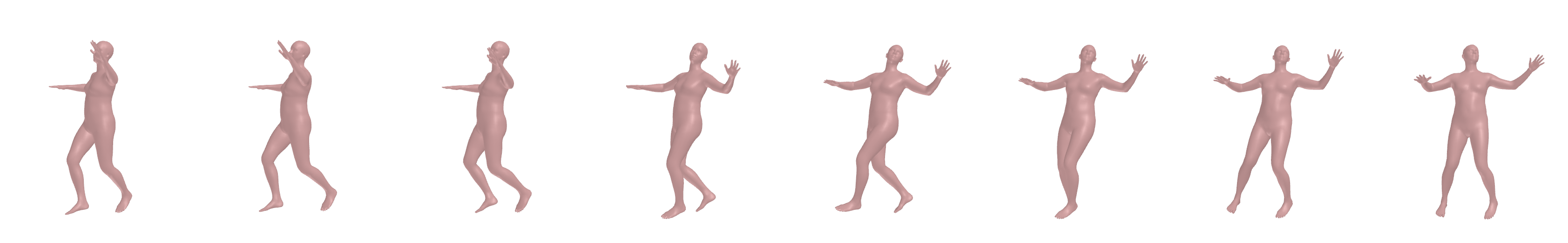}
\\ \hline
\end{tabular}
}
\caption{Visualization of sample generated tango dance sequences. Dance animations will be played in Adobe Acrobat Reader upon clicking.}
\label{fig:realism_results}
\end{figure*}

\subsection{Evaluation: Realism and Consistency} \label{sec:realism}

Sample generated dance sequences for tango are illustrated in Figure~\ref{fig:realism_results}. We observe that \cite{tang2018dance} tends to converge to mean pose and is lacking in dynamical variations. Comparatively, GAN methods including \cite{ren2020self} and our ablation experiment with WGAN demonstrates more variation in the range of dance motions, but often lacks in naturalness and appears unrealistic, especially in turning dance motions. Our MDOT-Net generates the most realistic looking dance motions which are also consistent with the tango style.

\noindent\textbf{User Study} \quad A single blind user study involving 8 dancers is engaged to judge the authenticity of the generated dance sequences and tabulated in Table~\ref{tab:realism}. For each dance style, 5 sample generated sequences of 20 seconds duration are ranked according to two criteria, namely 1) the naturalness and realism of the motion and 2) adherence to the specific dance style. The rankings echo our above qualitative observations, with the samples generated by our proposed approach being preferred over previous methods. This demonstrates the effectiveness of the optimal transport objective in generating realistic and consistent dance sequences.

\begin{table}[ht]
\resizebox{1\linewidth}{!}
{
\begin{tabular}{|l|c|c|c|}
\hline
Method & Realism & Consistency & FID \\ \hline
\cite{tang2018dance} & 5.9 & 5.8 & 105.5 $\pm 17.2$ \\ \hline
\cite{ren2020self} & 4.6 & 4.8 & 67.0 $\pm 9.3$ \\ \hline
\cite{huang2021dance} & 3.3 & 3.1 & 43.4 $\pm 7.5$ \\ \hline
Ablation: WGAN & 3.8 & 4.0 & 49.3 $\pm 5.2$ \\ \hline
Ablation: Remove GW & 2.3 & 2.1 & 32.4 $\pm 4.8$ \\ \hline
MDOT-Net (Ours) & \textbf{1.2} & \textbf{1.3} & \textbf{25.6} $\pm 3.3$ \\ \hline
\end{tabular}}
\caption{Results for Realism and Consistency}
\label{tab:realism}
\end{table}

\noindent\textbf{FID} \quad We randomly sample data dance sequences ranging from 100 frames to 250 frames and employ a pre-trained AGCN \cite{shi2019two} for feature extraction to evaluate the Fr\'{e}chet Inception Distance \cite{heusel2017gans}. The better FID result suggests that the optimal transport objective is more effective than existing methods and the WGAN in matching the model distribution with data distribution. This could be explained by the optimal transport objective not being prone to instability issues of adversarial training.

\begin{table}[h]
\resizebox{1\linewidth}{!}{
\begin{tabular}{|l|c|c|c|c|c|c|}
\hline
 &  & \multicolumn{5}{c|}{Multimodality} \\ \cline{3-7} 
\multirow{-2}{*}{Method} & \multirow{-2}{*}{Diversity} & Rumba & Cha Cha & Tango & Waltz & Average \\ \hline
Ground Truth & 63.5 & - & - & - & - & - \\ \hline
\cite{tang2018dance} & 18.2 & 13.2 & 16.7 & 14.8 & 10.7 & 13.9 \\ \hline
\cite{ren2020self} & 35.4 & 32.1 & 30.5 & 27.4 & 25.9 & 29.0 \\ \hline
\cite{huang2021dance} & 34.7 & 28.9 & 28.5 & 31.9 & 34.4 & 30.9 \\ \hline
Ablation: WGAN & 41.5 & 38.9 & 46.5 & 33.5 & 35.4 & 38.6 \\ \hline
Ablation: Remove GW & 58.7 & 52.3 & 55.4 & 46.8 & \textbf{50.7} & 51.3 \\ \hline
MDOT-Net (Ours) & \textbf{60.1} & \textbf{55.6} & \textbf{59.7} & \textbf{49.4} & 48.8 & \textbf{53.4} \\ \hline
\end{tabular}}
\caption{Results for Diversity (dance generated via different music) and Multimodality (dance generated on the same music)} 
\label{tab:diversity}
\end{table}

\subsection{Evaluation: Diversity and Multimodality} \label{sec:diversity}
We adopt the terminology of \cite{lee2019dancing}: \emph{diversity} refers to variations over the entire ensemble of dances generated from different music inputs whereas \emph{multimodality} pertains to dances generated from the same music. For quantitative evaluation, we employ a perceptual similarity metric \cite{zhang2018unreasonable}. For diversity, we generate 20 dance sequences conditioned on different music inputs. For multimodality, 5 dance sequences are conditioned on the same music. The pairwise feature distance is evaluated for each collection and experiments are averaged over 20 independent trials. The results reported in Table~\ref{tab:diversity} demonstrate significant improved diversity and multimodality over existing methods and the ablation setting of WGAN, indicating that MDOT-Net is more effective in preventing mode collapse.

\subsection{Evaluation: Cohesion and Unity with Music} \label{sec:music}
\begin{table}[t]
\resizebox{1\linewidth}{!}{
\begin{tabular}{|l|c|c|c|c|c|c|}
\hline
& \multicolumn{4}{c|}{Beats Matching (\%)} & \multicolumn{2}{c|}{Music-Dance Matching Ranking} \\ \cline{2-7} 
\multirow{-2}{*}{Method} & Rumba & Cha Cha & Tango & Waltz & \begin{tabular}[c]{@{}l@{}} Music from Dataset \end{tabular} & New Music \\ \hline
Ground Truth & 65.1 & 68.4 & 62.4 &  72.3 & 1.2 & - \\ \hline
\cite{tang2018dance} & 13.7 & 11.2 & 14.2 & 16.9 & 6.9 & 5.9 \\ \hline
\cite{ren2020self} & 44.7 & 40.2 & 39.9 & 50.7 & 5.6 & 4.7 \\ \hline
\cite{huang2021dance} & 52.3 & 46.5 & 52.6 & 58.2 & 3.6 & 1.9 \\ \hline
Ablation: WGAN & 49.8 & 49.4 & 55.7 & 58.8 & 3.0 & 3.8 \\ \hline
Ablation: Remove GW & 54.3 & 60.2 & 60.5 & 64.3 & 3.8 & 2.8 \\ \hline
MDOT-Net (Ours) & \textbf{63.2} & \textbf{65.4} & \textbf{64.7} & \textbf{73.1} & \textbf{1.9} & \textbf{1.2} \\ \hline
\end{tabular}}
\caption{Results for beats matching and user preference}
\label{tab:music}
\end{table}

\noindent\textbf{Beats Matching} \quad We evaluate the consistency of rhythmic articulation through beats matching. A dance beat is a local minimum in the mean joint speeds and it matches a music beat if the two events are within $\pm2$ frames. The beats matching ratio measures matching beats against total dance beats.

\noindent\textbf{Music-Dance Harmony} \quad Evaluating the music and dance coherence is a rather subjective task and we again engage an user study. 10 dance sequences (20 sec duration) are ranked by 8 users according to the perceived harmony with music. The consistent preference for MDOT-Net justify the effectiveness of the Gromov-Wasserstein distance.

\section{Conclusion}
In this work, we propose a MDOT-Net framework for generating 3D dance sequences conditioned on music. Through an optimal transport objective for matching the model and data dance distributions as well as a Gromov-Wasserstein objective for aligning the music and dance, MDOT-Net proves capable of generating realistic dances that are consistent with the dance style, display diversity and match the music. Extensive experiments demonstrate the effectiveness of the optimal transport and Gromov-Wasserstein objectives. For future work, we will generalize this cross-domain sequence-to-sequence generation framework to more applications.

\section*{Acknowledgements}
This work is supported in part by the Ministry of Education Academic Research Fund Tier-1 Project Grant (RG94/20) in Singapore, and in part by the NSERC Discovery Grant (RGPIN-2019-04575) and the UAHJIC Grants in Canada.

\clearpage
\small{
\bibliographystyle{ijcai22}
\bibliography{References}
}

\end{document}